\documentclass [12pt,twoside]{article}           
\usepackage[left,modulo]{lineno}
\usepackage{setspace}
\usepackage{textcomp}  

\countdef\arxiv=201

\ifnum\arxiv=0 \input cpreb.tex \fi

\ifnum\arxiv=1

\usepackage{epsfig,times,lscape}
\usepackage[usenames]{color}

    \ifnum\count102=1

    \fi

    \ifnum\count102=2
\fi

\newcommand{\nc}{\newcommand}

     \ifnum\count101=1
\nc{\qI}[1]{\section{{#1}}}
\nc{\qA}[1]{\subsection{{#1}}}
\nc{\qun}[1]{\subsubsection{{#1}}}
\nc{\qa}[1]{\paragraph{{#1}}}

\def\qpar{\vskip 2mm plus 0.2mm minus 0.2mm}
\def\qL{\hfill \break}
     \fi 

      \ifnum\count101=2
 \nc{\qI}[1]{\parindent=0mm \vskip 8mm 
{\centerline{\LARGE \color{red}#1}}\vskip 3mm}
\nc{\qA}[1]{\vskip 2.5mm \noindent 
{{\bf\large\color{blue}  #1}} \vskip 1mm \parindent=0mm}
 \nc{\qun}[1]{\vskip 1mm \noindent {\sl #1 }\quad }

\def\qL{\hfill \break}
\def\qpar{\vskip 2mm plus 0.2mm minus 0.2mm}

      \fi

\def\qth{\vrule height 12pt depth 0pt width 0pt}
\def\qtb{\vrule height 0pt depth 5pt width 0pt}

\nc{\qfoot}[1]{\footnote{{#1}}}

      \ifnum\count101=1
\def\qbu{\hfill \par \hskip 6mm $ \bullet $ \hskip 2mm}
\def\qee#1{\hfill \par \hskip 6mm (#1) \hskip 2 mm}
      \fi
      \ifnum\count101=2
\def\qbu{\hfill \par \hskip 4mm $ \bullet $ \hskip 2mm}
\def\qee#1{\hfill \par \hskip 4mm (#1) \hskip 2 mm}
      \fi

\def\qparr{ \vskip 1.0mm plus 0.2mm minus 0.2mm \hangindent=10mm
\hangafter=1}

     \ifnum\count101=1 
 \def\qdec#1{\parindent=0mm\par {\leftskip=2cm {#1} \par}}
     \fi
     \ifnum\count101=2
  \def\qdec#1{\parindent=0mm \par {\leftskip=1cm {#1} \par}}
  
  \def\qcitb#1{\noindent \hbox to 102mm{\hfill \small #1} \vskip 1mm}
      \fi

 \def\qpages#1{\count102=0{\loop\advance\count102 by 1
 \null \vfill\eject \ifnum\count102<#1 \repeat}}

\def\qth{\vrule height 12pt depth 0pt width 0pt}
\def\qtb{\vrule height 0pt depth 5pt width 0pt}

\def\qv{\vskip 0.1mm plus 0.05mm minus 0.05mm}
\def\qhu{\hskip 0.6mm}
\def\qhv{\hskip 3mm}

\def\qhw{\hskip 1.5mm}
\def\qleg#1#2#3{\noindent {\bf \small #1\qhw}{\small #2\qhw}{\it \small #3}\qv }
\fi

\begin{document}
\thispagestyle{empty}



\markboth{{\sl \hfill  \hfill \protect\phantom{3}}}
        {{\protect\phantom{3}\sl \hfill  \hfill}}

\color{yellow} 
\hrule height 20mm depth 10mm width 170mm 
\color{black}
\vskip -1.8cm 

\centerline{\bf \Large Analytical history}
\vskip 2mm
\centerline{\bf \Large }
\vskip 15mm
\centerline{\large 
Bertrand M. Roehner$ ^1 $ }

\vskip 4mm
\large

\vskip 2mm
\centerline{\it \small Provisional. Version of 10 February 2017. 
Comments are welcome.}
\vskip 2mm

\vskip 2mm

{\normalsize 
1: Institute for Theoretical and High Energy Physics (LPTHE),
University Pierre and Marie Curie, Paris, France. 
Email: roehner@lpthe.jussieu.fr
}

\vskip 15mm

{\it The purpose of this note is to explain what is
``analytical history'', a modular and testable analysis of
historical events introduced in a book published
in 2002 (Roehner and Syme 2002). Broadly speaking,
it is a comparative methodology for
the analysis of historical events.
Comparison is the keystone and hallmark of science.
For instance, the extrasolar planets are
crucial for understanding our own solar
system. Until their discovery, 
astronomers could observe only one instance.
Single instances can be described but they cannot be understood
in a testable way. In other words,
if one accepts that, as many historians say, ``historical
events are unique'', then no testable understanding can
be developed.} 
\qpar

%
\begin{figure}[htb]
\centerline{\psfig{width=7cm,figure=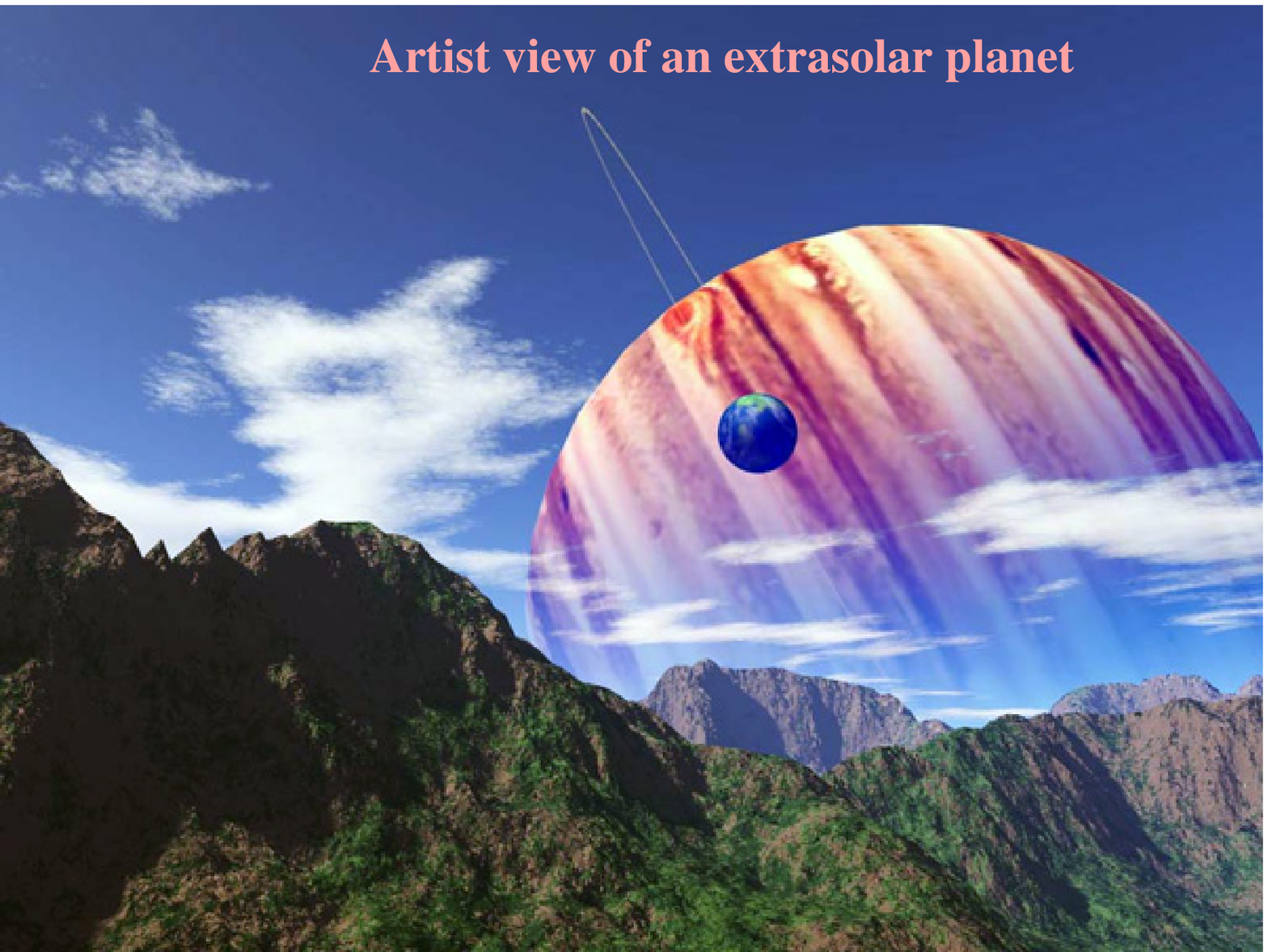}}
\qleg{Fig.\qhu 1\qhv Exoplanet.}
{Comparison is the linchpin of science. Knowing
the characteristics of extrasolar planetary systems
will give us a testable understanding of our own
solar system. In the same way, comparison with other
stars allows us to test our understanding of the Sun.
It is true that each exoplanet or each star is 
in its way unique but that should not preclude
comparison.}
{}
\end{figure}

{\it Because taking a comparative view is really the key of
the approach that we propose, it will be illustrated
in the following section through two examples.}

\qI{Manifest for viewing the world in comparative perspective}

The following case suggests that in human
affairs making comparisons does not come naturally.
It requires an effort.
It is easier to remain focused on the little world in
which one lives than to look outside.

\qA{How to manage street cleaning}

In the past two or three decades the streets of Paris
have become more and more dirty. Papers, 
empty cans, plastic bags,
cigarette buts litter the sidewalks. Actually,
even shortly after being cleaned 
the streets are nevertheless not really neat which shows
that there is a deep flaw in how cleaning is 
performed.
Although 
several mayors and city councils alternated they were
all unable to solve the problem in spite of the promises
made to their electors. Yet,
this is a challenge to which
all cities are confronted and which many are able to handle
successfully. Therefore the solution seems quite obvious.
What should one do?
One needs only to study
the management and techniques in use in successful cities such as
for instance Geneva or Munich and
then implement them in Paris. 
\qpar
The fact that this is {\it not}
done suggests that to adopt a comparative standpoint is
not something natural.
\qpar
Needless to say, this reflection is not limited 
to street cleaning.
As basically all cities are  confronted to the
same issues%
\qfoot{E.g. homeless people sleeping in the streets, 
selective waste sorting, 
air pollution, public transportation.}
a {\it close examination} of the solutions
that worked the best elsewhere should always be a 
mandatory first step ahead of any policy selection.
At least it would give a clear
understanding of the means that do work even if putting 
them into effect may, for a variety of reasons, represent
a challenge.
\qpar

In order to show that the adoption of a
comparative framework really makes a crucial difference
in many important issues we give a second illustration.

\qA{How to end secessionist movements and civil wars}

In French, the American ``Civil War'' is called 
the ``Guerre de S\'ecession'' (``Secession War''). 
Whereas not all
civil wars are caused by secessionist movements%
\qfoot{For instance the civil wars in Germany (1618--1648),
Russia (1917--1922),
Ireland (1922--1923), Spain (1936--1939), China
(1927-1949), Vietnam (1945--1975) were not primarely caused
by any secession.}%
,
it is true that separatist movements often lead to 
civil wars. In our time (2017) many civil wars are
under way%
\qfoot{E.g. in Afghanistan, Iraq, Libya, Myamar, the
Philippines, Somalia, Sudan, Syria, Yemen.}
and a majority of them  are caused by separatist
movements. Yet, is it not remarkable that when 
accounting and explaining such events American
media people seem to be completely oblivious of 
how the Civil War came to an end in their own
country? Even a cursory comparative view 
would tell us two things.
\qbu Civil wars are always bloody and ``dirty'' wars.
One only needs to recall ``Sherman's March to the Sea''
in Georgia at the end of 1864 to emphasize that the civilian
population suffers a lot in such wars%
\qfoot{It is claimed that Sherman's campaign
was conducted in agreement
with the ``Lieber Code'' (Instructions for the government
of Union armies in the field)
which had been 
approved by  President Abraham Lincoln in 1863. However,
to our best knowledge, even
today there is no reliable estimate of the number
of civilian casualties. Therefore it is difficult to know
how this total war campaign was really carried out.}%
.
\qbu The clearest effect of foreign intervention
is to make such wars drag on longer. The ``Thirty Year War'' in
Germany (1618--1648), the Civil War in China (1919--1949) or
the Vietnam War (1945--1975)
are cases in point.
\qpar

It is hoped that, despite their brevity, 
the previous indications
will convince readers that in many areas
adopting a comparative perspective is not just a
matter of
academic inclination but has in fact far 
reaching implications.
\qpar

It can also be observed that
experimental physics, which so far has been
the most successful of all scientific fields, 
is entirely based on comparing the results of 
experiments performed under different conditions.
In other words, for physicists the
comparative approach presented in this paper
should be quite natural. That was indeed the case
at the end of the 19th century but no longer
nowadays
for over the past century physics has become
more and more theoretical.

\qI{Analytical, modular and testable}

The purpose of the rest of this note is to describe the
comparative methodology developed in   
``Pattern and Repertoire in History'' (Roehner and Syme 2002).
\qpar

First of all a comment about the book's title may be useful.
It was not chosen by the authors but by the publisher in conjunction
with Prof. Charles Tilly who was one of the book's reviewers.
\qpar
How should the word ``repertoire'' be understood? ``Repertoire''
means a set of actions but here it refers
to a {\it limited} set of action. 
In reacting to an event a country (or more generally a group
of people) will pick up
a type of action in the limited repertoire of its past actions.
Most often the selected action is one that had already occurred
several times in the past or a mixture of two or three past 
actions.
\qpar
In other words, the emphasis here is on {\it recurrent} actions
which means that the repertoire will be limited to two
or at most three items. If, instead of 
being a very narrow repertoire, it would contain many items,
then the analysis would not have any predictive power whatsoever
and would be quite useless.
\qpar

The authors' suggestion for the book's title 
was ``Analytical History''
which is indeed the title of its first chapter. With the
benefit of hindsight we are not sure it would have been a 
better title. 
\qpar

What is the main idea behind these notions of pattern and
recurrent events?
\qpar

Let us assume we are interested in separatist movements of
a given type. If we can collect historical evidence about
100 such episodes and all of them failed, then there is
a strong likelihood that the 101st which we see developing
under our eyes, will fail also. 
\qpar

This, in a nutshell, is the approach of
analytical history: it relies on
studying sharply defined events on a large number of cases. 
Not surprisingly, its
practical implementation raises a number of questions.
\qpar
We will see that in fact there are three successive
steps (Fig. 2). 
\qee{i}Decomposing complicated events into simple
modules. 
\qee{ii} Collecting evidence about a large
number of cases involving these modules.
\qee{iii} Proposing testable predictions.
\qpar
These steps broadly correspond to the three expressions used
in the title of this section: {\it analytical} refers
to the decomposition process, {\it modular} refers to
the study of simple modules on a large number of cases,
{\it testable} refers to the fact that the
previous investigation must be confronted to {\it new} evidence
in order to check whether or not it is confirmed by observation. 
\qpar
The last step
is essential as explained in the next subsection.

%
\begin{figure}[htb]
\centerline{\psfig{width=15cm,figure=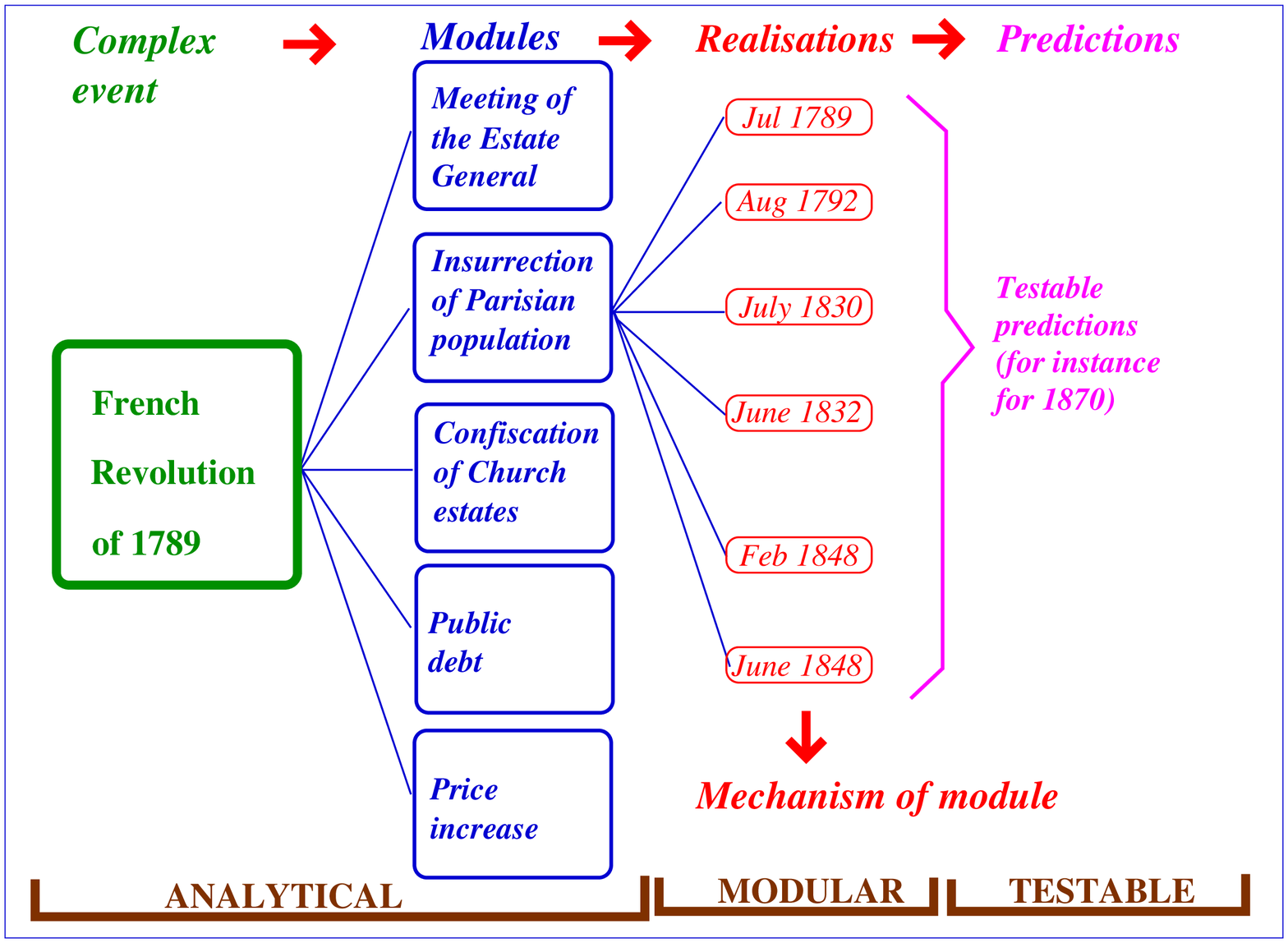}}
\qleg{Fig.\qhu 2\qhv Presentation
of the successive steps in investigating complex events.}
{First any complex event is decomposed into simpler modules
which will then be studied separately. For each module
the second step is to collect historical evidence in
the form of
a sample of realisations. Then, these realisations must
be studied comparatively in order to find which
rules they obey. These rules will then lead to testable
predictions. If such predictions are confirmed
by observation the proposed
mechanism will be conforted.}
{}
\end{figure}

\qA{Testable predictions}

In a general way in the social sciences
for any observation one can 
propose an infinite number of explanations.
None may be really wrong for in fact each one will 
just address a different facet of the observed effect.
Similarly,
it is not wrong to say that the movements of a pendulum
are under the influence of the gravitational field
of the Sun and Moon. However, this effect accounts for
only a minute fraction of the actual movement; thus, this
effect cannot be seen unless one can perform high accuracy 
measurements.
\qpar
How then can one distinguish between main factors
and accessory factors?
The answer is very simple.
\qpar
Each ``model'' (whether mathematical or not)
should be, not just a closed explanation, but
at the same time should provide {\it testable predictions}.
If in the course of time
the predictions turn out to be correct there will be 
a good likelihood that the ``model'' is able to capture
the bulk of the phenomenon under consideration.

\qA{The sunrise paradigm}

Why did we write the word ``model'' within quotes?
The reason again is fairly simple. \qL
For most economists and
economic historians the word ``model''  refers to a mathematical
model. Actually, in order to make predictions
there is no need for a mathematical model. This can
be seen easily through what may be called the sunrise paradigm.
\qpar

Suppose that on a given day, let us say Monday, 
you have observed that the sun rises at 8:36 am. Then, on
Tuesday and Wednesday you see it rise at 8:37 and 8:38 respectively.
You will of course be tempted to predict that on Thursday
it will rise at 8:39, a prediction indeed confirmed by
observation. Emboldened by this success, you may feel assured
that on Friday sunrise will happen at 8:40. Alas, it occurs
also at 8:39. This may suggest that to give the times in minutes
is perhaps not accurate enough. Data expressed in minutes and seconds
are as follows (from Monday to Friday)%
\qfoot{These are real data for Paris between 12 December
2016 (Monday) and 16 December (Friday). The source
is the website http://sunsetsunrise.com.}%
:
8:36:45, 8:37:37, 8:38:27, 8:39:14, 8:39:58. For these
data the day-to-day changes, namely 52s, 50s, 47s, 44s, 42s
display indeed a fairly regular pattern.
In short, once one has a regular pattern, one can 
propose predictions. There is no need of a mathematical
model. Actually, for the sunrise example astronomy
does not offer any mathematical model whatsoever
because the rotation of the Earth
around its axis in 24 hours is a purely empirical fact
(contrary to the rotation time around the Sun which
is controlled by the distance to the Sun).
\qpar

In all sciences the first significant steps were the
discovery of patterns. For instance
the three Kepler laws which rule the orbits of planets
can be seen as a pattern. It was of utmost importance
because it opened the road to Newtonian mechanics.
\qpar

Readers may think that the task of finding patterns
was easier for planets than for socio-economic
events because the later are much more diverse.
It is quite right that defining and setting up
an {\it homogeneous} set of events is a major challenge.
However, as explained in the next subsection,
this difficulty is also present in other fields
such as medicine.

\qA{Homogeneous sets of events: the paradigm of medical diagnosis}

When you visit your doctor because (for instance) you feel
stomach pains, she will ask you a number of questions.
Are you feverish? Do you feel pain immediately after your
meals? And so on. Then, based on your answers,
your doctor will try to
establish a connection with
some recognized pattern of symptoms.
May be, she will be led to the conclusion that
it is a gastroenteritis, that is to say
a viral infection. \qL
This example has several
similarities with the task of defining an homogeneous
set of events.
\qbu The symptoms of two persons suffering from
gastroenteritis are not exactly the same. Depending upon
their age, their diet, the season there will be
differences. Whether these differences are of significance
or merely accessory is a difficult decision.
Each patient in some way is unique and to some extent
each illness is unique too. That is why, not infrequently
starting from the same set of symptoms, two doctors 
may draw different conclusions.
\qbu In order to reduce the uncertainty one
must get additional evidence for instance
by performing a blood test.
\qbu The ultimate test for deciding whether the diagnosis
was correct is to see whether the treatment is able to cure
the patient. This is the analogue of the successful
predictions mentioned above.
\qpar

At the source of each illness there is usually a
``mechanism'': viral infection, cancerous cells
and so on. A well-defined mechanism (e.g.
infection by an adenovirus) will produce fairly
reproducible symptoms. Similarly, focusing on
a well-defined mechanism is a good way to create
a fairly homogeneous sample of events. 
Examples are given in the next section.

\qA{Analytical history}

So far our explanations concerned modular history. 
Before we close this introduction we must explain the
purpose of the analytical procedure. It is obvious that
major historical events are unique. There has been only
one French Revolution of 1789 and only one
American Revolutionary War. However, such events have
many facets which it is possible to study 
separately. This process is schematized in Fig. 2.
It is in fact similar to what is done in chemistry when
a mixture is separated into its components, i.e.
the chemical compounds which compose it. 
For instance, whiskey is a mixture mostly of water and
alcohol.

\qI{Examples of mechanisms}

What we call a mechanism is what Vilfredo Pareto called
the ``residue'' of a class of events (see below).
This notion 
plays a key role because one (or several) mechanisms are
associated with each class of events.
In this section we give
several examples of mechanisms.
The first corresponds to the case of Fig. 2; the second
and third
belong to economic history; the fourth is about
separatist movements. Finally, the last
concerns the history of the American Revolutionary War.

\qA{Insurrections of the Parisian population}

The insurrections of the Parisian population were an
important element of the French Revolution. 
The one which took place
on Bastille day is well known but there were
many others some of which are listed in Fig. 2.
Most of them followed the same scenario. Thus,
it makes sense to analyze a sample of such
events. Just as an illustration of the
conclusions that can be derived, it turns out
that a key-element in the success or failure
was the attitude of the ``National Guard''.
This was a military unit whose men and officers
were ordinary citizens. Whenever the National
Guard sided with the rebels the insurrection
was successful. Clearly, this rule can be used
for making predictions.

\qA{Economic growth by sector}

Between 1980 and 2014 the Chinese Gross Domestic Product (GDP)
per employee in the primary, 
secondary and tertiary sectors (let us denote
these variables by $ g_1,  g_2, g_3 $)
all increased exponentially and in fairly parallel ways.
That is why their ratios remained comprised within
fairly narrow intervals. 
On average: $ g_2/g_1 =5.5,\ g_3/g_2=0.75 $.
Thus, an expanding secondary sector will increase the global
GDP per employee $ g $ whereas an extension of the tertiary
sector will decrease it (if at the same time $ g_2/g_1 $ remains
constant). This is a very simple mechanism but it can
explain both the rapid development of China and the leveling off
of wage increases in developed countries that was
observed in past decades%
\qfoot{In Baaquie et al. (2016), in analogy with the 
well-known demographic transition, this effect was called a
``wage transition''.}%
.
\qpar
If one assumes that in its development China will experience
the same changes of sectorial weights as industrialized
countries, then one can propose a long-term
prediction for the evolution of $ g $. This is what
was done in Baaquie et al. (2016).

\qA{Real estate speculative bubbles}

When real estate prices go up markedly
there is a critical level above which people who wish to
buy for their own need (let us call them ``users'') can
no longer afford to buy. Above this threshold, the market
is more and more dominated by investors who buy
with the objective of selling at a higher price 
some time later. However, as prices continue to
climb, less and less investors are willing to buy 
for the following reason.\qL
Usually during such a speculative episode
prices increase
faster than rents. As a result the yield of capital
invested in housing falls. The first sign of
a coming crash is the contraction of the
transaction volume as investors become
less eager to buy. Not surprisingly, this
contraction is particularly severe 
in the ``buy-to-let'' market.
\qpar

This mechanism can be seen at work in many speculative
episodes. As a matter of fact observation shows that
during such episodes prices display almost always the
same pattern of changes. First over a period
of time that may range from 5 to 10 years, there is
a multiplication of
real prices by a factor 2 or 3; this is then followed
by a downturn and a period of falling prices which
is usually of same length as (or slightly shorter than) 
the upward phase.
The existence of such a pattern opens the possibility
to make predictions.
\qpar

One such prediction was made 
in September 2003 in a reply to a review of ``Pattern
and Repertoire'' (Roehner 2003a):
\qdec{``Some economists claim that demand is strong
and will remain so; others contend that there may be
a small correction in the next two years. But surprisingly
few researchers have tried to take a close 
look at former price peaks to better understand what may be 
in store. Yet people in Boston, New York or San Francisco
may remember that 13 years ago the bust of a
housing bubble brought about financial troubles for
many saving institutions. Will the same phenomenon
repeat itself in the 5 or 6 years to come?}
Two subsequent papers (Roehner 2003b, 2006)
gave a prediction for how
prices will fall in California which turned out
to be fairly correct (see Richmond et al. 2013).

\qA{Separatist movements}

``Pattern and Repertoire'' was published together with
a companion volume entitled ``Separatism and Integration''
(Roehner and Rahilly 2002, a more recent presentation
can be found in Roehner and Rahilly 2016),
Whereas ``Pattern and Repertoire'' gives several
short (and mostly qualitative) descriptions of recurrent events, 
``Separatism and Integration'' proposes a thorough analysis
of separatist movements. It turns out that the key-factors
in such movements are always very much the same.
For instance, they originate mostly in islands or 
in remote corners of a country. Moreover, historical
roots are a crucial element. These connections can
be expressed in a quantitative way in the forms of
graphs (see Ch. 8 of Roehner et al, 2002).

\qA{The ordeal of loyalists}

Consider a war between two sides $ A $ and $ B $.
Let us further assume that a city $ a $ belonging to side
$ A $ is about to be invaded by the army of $ B $. In
addition let us suppose that among the citizens of $ A $ 
some (let us call them loyalists)
who openly showed their preference for 
$ B $ have been put in prison. Obviously, if left
in prison they will be freed as soon as the forces of $ B $
will have overtaken the city. 
Then they will probably join
the army of $ B $. This is clearly an 
outcome that any responsible
commander of the $ A $ forces should try to avoid. 
There are not many solutions. The prisoners must 
be either transported away or eliminated that is to
say executed. 
\qpar
Numerous examples of this scenario can be observed
in the history of all countries who experienced
this kind of conflict. As it follows almost always the
same rules one can use them to make
predictions. For instance, during the ``American
Revolutionary War'' every time an area was under the 
imminent threat
of being invaded by the British one would expect the 
loyalists to be rounded up (if they were not in prison
already), tried and possibly executed. Three cases
of this kind are mentioned in Roehner (2003a).
\qpar
Due to the fact that
during the protracted six-year long war of independence
there was no clearly defined front line, there  
have probably been other cases in which areas previously
under American control were invaded by British troops.
The previous rule suggests to give a close look at the
treatment of Loyalists under these circumstances.
This is an interesting case because it shows that the
predictions do not necessarily need to be about the
future, they can also help us to identify past events
which so far did not attract much attention.

\qI{Roots of modular history and previous attempts}

Almost all reviews of ``Pattern and Repertoire'' observed
that this was not the first attempt to make the study of history
more ``scientific''. The undeclared implication was that
this attempt will probably fail just as previous ones
have failed. Yet, one can be fairly sure that some time in
the future this objective will indeed be realized. Why?
\qpar
One possible answer can be given by
considering the field of astronomy (Fig. 3).
%
\begin{figure}[htb]
\centerline{\psfig{width=15cm,figure=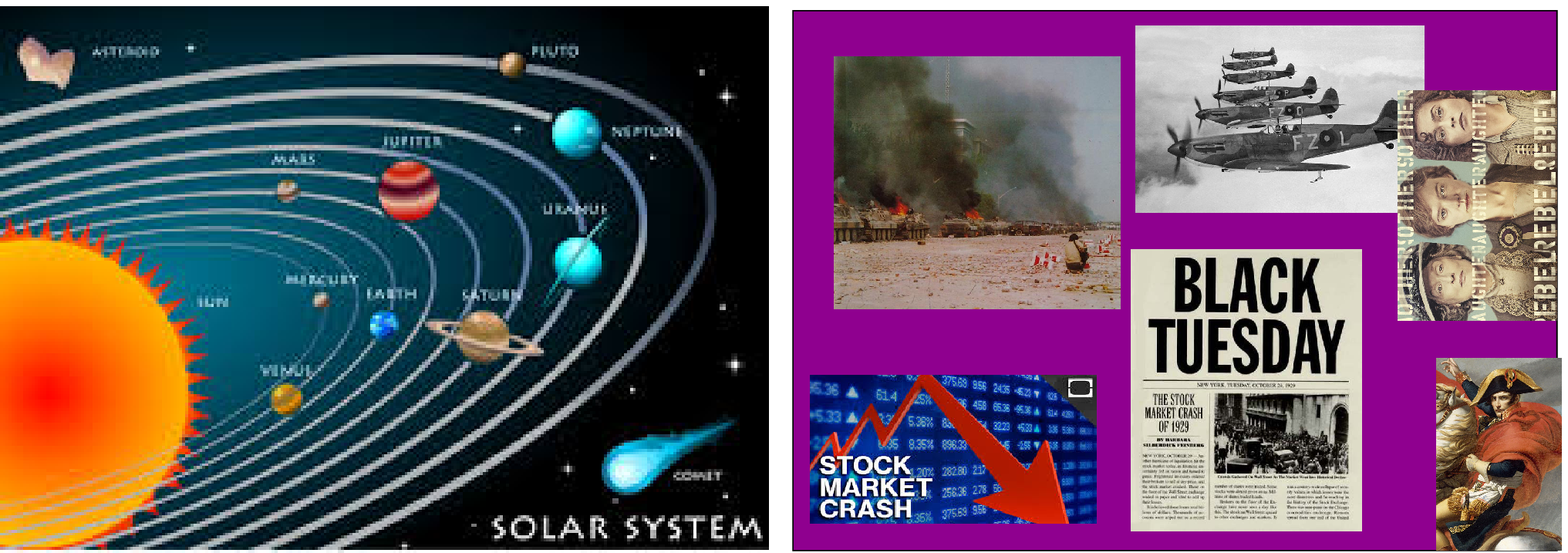}}
\qleg{Fig.\qhu 3\qhv Observation in astronomy and history.}
{For thousands of years people have recorded
the trajectories of
planets without being able to make sense of their
observations. Similarly for thousands of years people
have recorded historical events without being able
to make much sense of them. Although from planets, to
asteroids, to meteorites, to comets the solar system
is far from being simple, it is true that there is an
even greater diversity in human events. That is why it is
a good idea to
start by studying classes of similar events.}
{}
\end{figure}

Like astronomy, history has for first objective to make observations
and to do that as accurately as possible. Together with
its parent field astrophysics, astronomy also wants to organize and
analyze its data in ways which ``make sense''. Although astronomical
observations started thousands of years ago, astronomy became a real
science only a few centuries ago. 
\qpar
Our objective is to raise the status of history
from facts collecting to being capable of providing testable
predictions. Countless events have been recorded in numerous countries
and over many centuries. Our goal now is to make sense of them.
\qpar
The transformation of astronomy into a science
was not achieved all
of a sudden but rather gradually through a succession of steps.
The names of 
Eratosthenes of Cyrene (-276 to -195),
Hipparchus (-190 to -120), Ptolemy (100 to 170) come immediately
to mind. It would be a mistake to think that these persons
were all ``real'' scientists. For instance, one the three major
books written by Ptolemy was about astrology. Even Johannes Kepler
wrote three books on astrology.
\qpar

In short, it is stepwise that astronomy became the 
science that we know presently. Similarly, the transition from
purely descriptive history to analytical history will come
by degree. That is why in the following sections we recall
some of the important landmarks on this long road.
What matters is to go forward not backward. We hope that
these previous achievements will give a better sense of
the direction in which one needs to progress.

\qA{Pareto: emphasis on the core of a set of events}

Vildredo Pareto was a great economist but also an important
sociologist. It is unfortunate that his major work (Pareto 1917)
attracted little attention and is now largely forgotten.
In this work, Pareto develops an approach which is similar
to ours in the sense that it is based on the analysis
of samples of {\it similar events}. For instance,
speaking of (mostly magical) methods 
used by people to keep away storms, Pareto says (section 865):
\qdec{``Would we know only one of these cases it would be impossible
to infer anything from it. On the contrary, by analyzing
a large sample of such events we can identify the 
invariant kernel of such actions'' (my translation
from the French version).}

As can be seen from this example, Pareto's objective is
very ambitious because he includes in his
analysis not only historical events but also actions dictated
by superstition, tradition or religion. In other words,
is is not only about facts which are well defined in time
and space but also about actions described (in fairly
loose ways) by commentators%
\qfoot{Most of the sources used by Pareto are works
by ancient Greek and Roman writers. The fact
that in the 21st century
few persons are familiar with these authors is of course
an obstacle. Another is the sheer size of Pareto's book,
nearly 1,800 pages; the English edition comprises 4 volumes.}%
.
\qpar
Table 1 lists the notions introduced by
Pareto and the equivalent notions used 
in ``Pattern and Repertoire''
\qpar

\begin{table}[htb]

\small

\centerline{\bf Table 1\quad Equivalence table between
the notions introduced by Pareto and those used in analytical
history}

\vskip 5mm
\hrule
\vskip 0.7mm
\hrule
\vskip 2mm

$$ \matrix{
\qtb
\hbox{Pareto} &  & \hbox{Analytical history} \cr
\noalign{\hrule}
\qth
\hbox{Logico-experimental science} \hfill & \longleftrightarrow & 
\hbox{Testable science} \hfill\cr
\hbox{{\color{red} Residue}} \hfill& \longleftrightarrow & 
\hbox{{\color{red} Mechanism} (derived from a set of realizations)} \hfill\cr
\hbox{Derivative} \hfill& \longleftrightarrow & 
\hbox{A realization (of a given module)} \hfill\cr
\hbox{Derivations} \hfill& \longleftrightarrow & 
\hbox{Accessory features (in a realization of a module)} \hfill\cr
\hbox{\it Decomposition equation} \hfill & & \cr
\qtb
\hbox{\normalsize \it derivative = {\color{red} residue} + derivations}
\hfill & \longleftrightarrow & 
\hbox{\normalsize \it realization = {\color{red} mechanism} +
  accessory features} \hfill\cr
\noalign{\hrule}
} $$
\vskip 1.5mm
\small
{\it Source: Pareto 1917, sections 865--867 (p. 458--459 in the
French edition).}
\vskip 5mm
\hrule
\vskip 0.7mm
\hrule
\end{table}

At first sight the choice by Pareto of the word ``residue''
to designate the mechanism of a module may surprise. 
In fact, he uses
this word in the sense that it has in chemistry: when one filters
a solution containing a suspension, the residue
which remains in the filter is usually the most important component.

\qA{Marc Bloch: emphasis on sharply focused comparative studies}

In 1924 the French historian Marc Bloch (1886-1944) published
a book which was a landmark in the development of comparative
history for at least two reasons.
\qbu Instead of being a broad comparison, the study was focused
on a sharply defined phenomenon, namely the assumed ability of
kings to cure diseases.
\qbu This phenomenon was studied comparatively in two countries,
France and Britain, over a period of several centuries. Moreover
it was studied not only from the point of view of history but also
from a sociological perspective.
\qpar

What we propose in ``Pattern and Repertoire'' parallels these
features. Of course, one may regret that the author examined
only two cases. We are told that Marc Bloch knew some 10 languages
which would have enabled him to consider other countries, for instance
Austria, Hungary, Poland, Spain or Sweden. The existence or
absence of a similar tradition and belief in such countries
would have given more precise information about the key-factors
which control this phenomenon.

\count101=0  \ifnum\count101=1

Scrofula was therefore also known as the King's Evil. From 1633, the
Book of Common Prayer of the Anglican Church contained a ceremony for
this, and it was traditional for the monarch (king or queen) to
present to the touched person a coin – usually an Angel, a gold coin
the value of which varied from about 6 shillings to about 10
shillings. In England this practice continued until the early 18th
century
King George I put an end to the practice as being "too
Catholic".[citation needed] The kings of France continued the custom
until Louis XV stopped it in the 18th century, though it was briefly
revived by Charles X in 1825.

\fi

\qA{Charles Tilly: emphasis on large $ n $ investigations}

Back in 2002
when ``Pattern and Repertoire'' was published
Prof. Charles Tilly (1929--2008) predicted that
for historiography the book would be a sort of tornado.
That did not happen, however.
\qpar

Charles Tilly was in his time a very influential
political scientist and historian.
In his book ``Popular Contention in Great Britain, 1758--1834''
Tilly collected and analyzed information about
over 8,000 ``contentious gatherings''. 
This was a major step forward and
a model for the investigations of large $ n $ samples.

\qA{Stanley Lieberson: emphasis on cumulative knowledge}

In a book (Lieberson 1985) and an article (Lieberson 1992)
Harvard sociologist Stanley Lieberson underlines that 
in the social sciences the main challenge is to
achieve development in a {\it cumulative way}.
This is indeed essential for otherwise
there will be a number of bright spots
but without connection with each other. That does
not make a science.
\qpar
What made the success of physics is that 
there was a cumulative development and a
synergy between its different
fields. This did not came by chance.
Table 2 shows that over more than three centuries
with a determination and a patience which in
retrospect seems quite remarkable. 
experimental physicists improved the accuracy of
their measurement of the speed of light.

\begin{table}[htb]

\small

\centerline{\bf Table 2\quad Successive measurements 
over three centuries of the
speed of light in vacuum.}

\vskip 5mm
\hrule
\vskip 0.7mm
\hrule
\vskip 2mm

$$ \matrix{
\hbox{Year}\hfill & \hbox{Measurement} & & \hfill \hbox{Error bar}
& \hbox{Accuracy} \cr
\qtb
\hbox{}\hfill & \hbox{[meter/second]} & & \hfill \hbox{[meter/second]}
& \hbox{[per thousand]}\cr
\noalign{\hrule}
\qth
1675 &	220,000,000 & \pm & \hfill 100,000,000 & 500\ \hbox{\textperthousand}\cr
1862 &	298,000,000 & \pm & \hfill   500,000 & 1.7\ \hbox{\textperthousand}\cr
1907 &	299,710,000 & \pm &  \hfill   30,000 & 0.10\ \hbox{\textperthousand}\cr
1926 &	299,796,000 & \pm &  \hfill    4,000 & 0.012\ \hbox{\textperthousand}\cr
1950 &	299,792,500 & \pm &   \hfill   3,000 &0.010\ \hbox{\textperthousand} \cr
1958 &	299,792,500 & \pm & \hfill    100 &0.0003\ \hbox{\textperthousand}\cr
\qtb
1972 &	299,792,456 & \pm &\hfill 1.1& 3\times
10^{-6}\ \hbox{\textperthousand}\cr
\noalign{\hrule}
} $$
\vskip 1.5mm
\small
Notes: The speed of light in air is smaller than
in vacuum by 90,000 m/s. This does not mean that it {\it must}
necessarily be measured in vacuum for, knowing the index
of refraction of air (namely 1.0003)
it is easy to derive the speed in vacuum from a measurement
performed in air. The accuracy of the measurement is defined as the
error bar divided by the measurement value.
{\it Source: Wikipedia article entitled ``Speed of light''.
This list is certainly not complete for
other measurements were probably made between 1675 and 1862.}
\vskip 5mm
\hrule
\vskip 0.7mm
\hrule
\end{table}

This contrast with what we see in history. 
As already mentioned after Marc Bloch's study of the
assumed ability of kings to cure diseases which focused
on Britain and France, nobody tried to extend this
investigation to other countries. Instead, for instance
in France, historians moved to other topics which seemed
more fashionable. This was the so called ``New History''
which focused mostly on cultural history.

\qA{Development of comparative data bases}

In the last fifteen years 2000--2015, especially
in the United States, there has been a rapid development of
data bases set up in a comparative perspective.
We will mention particularly two of them.
\qbu IPUMS
(Integrated Public Use Microdata Series) data base developed
by the ``Population Center'' of the University of Minnesota
provide individual census records for all US
censuses and for many censuses of a sample of 
foreign countries which is progressively expanded. 
In the IPUMS label it is the word ``Integrated'' which is 
the most important. Indeed, before new data are included
care is taken to ensure their compatibility with existing
data.
\qbu From its very beginning the project of Wikipedia was
to be a multinational database. This was conducive to
the adoption of a comparative perspective. 
At present time (2016)
many of the non-English articles are translations 
or summaries of the relevant article in English 
but one can hope that in following decades 
the contribution of other countries will increase 
so that eventually it
becomes truly multinational.
In its last chapter ``Pattern and Repertoire'' presented the
project of a ``Very Large Chronicle'' (VLC) which would
be a systematic data source of historical events. Wikipedia
was a step in this direction. However, in the VLC project
each event would be connected to a set of primary archive
sources. This is a fairly ambitious project but is probably
the only way to ensure reasonable reliability.

\qA{Thomas Piketty: revival of comparative analysis}

Thomas Piketty's book ``Capital in the 21st century''
has had a tremendous success in the United States.
Some 200,000 copies were sold (Tracy 2014).
However, almost none of its numerous reviews 
pointed out what can be seen as a key-aspect of
the investigation, namely that it was a comparative
study both in time (longitudinal analysis) and
across nations (transversal analysis).
\qpar
With regard to comparison in time
it is a common saying that for
most economists the world starts in 1945. 
On the contrary, Thomas Piketty's study goes 
back to the mid-19th century. 
\qpar

At first sight the fact that his study extends to
several industrialized countries may not seem
very impressive because some econometric studies
use samples of over hundred countries.
In such cases, however, the data that they use 
are readily available for instance on the websites
of the ``World Bank'' or ``International
Monetary Fund''. On the contrary, for each of the cases
he considered, Prof. Piketty had to develop 
specific historical datasets. This is uncommon.
\qpar
Does it mean that there will be a renewal
of comparative studies? Although we would be happy
to answer ``yes'' a lucid assessment raises
many doubts. Just as one swallow does not make 
a summer.

\qI{Conclusion} 

Some historians do not agree with the proposal
that there can be laws in history.
Well, in sociology there are some laws which do have
a broad validity; why should it be different
in history?
\qpar
Perhaps the reader would like us to give an example
of such a broad law for it is true that there are not many. 
\qpar
Detailed death statistics by cause of death became available
in most industrialized countries from the mid-nineteenth century on.
Similarities across countries were soon identified.
One of them is the impact of marital
status on age-specific death rates (Bertillon 1879).
The rates for married people were found to be 
two or three times lower
than for single, divorced or widowed persons.
For death by suicide the ratios were found
to be even higher (Durkheim 1897).
To our best knowledge (Richmond et al. 2016)
the validity of this law extends
to all time intervals and all countries (including
an Asian country such as China)
for which reliable data are available. 
\qpar

In recent years Prof. Peter Turchin developed {\it Cliodynamics};
what is its difference with analytical history? Broadly
speaking, it can be said that cliodynamics wishes to 
develop mathematical models (somehow as in econometrics)
whereas analytical history
uses the approach of {\it experimental} science. 
That does not exclude building models but only once
the mechanism has been well understood thanks to
well focused observations.
\qL
It would be pointless to discuss which approach is better.
Actually, all attempts of this kind are welcome.
There should be a profusion of them just as there are
numerous software companies in the 
Silicon Valley. Then, each
approach should be judged on its fruits, that is to say the
number of successful predictions that it can offer.
In this way the best options should emerge.
\qpar

A last word may be useful which concerns non-human societies.
In ``Pattern and Repertoire'' the chapter about wars for
territorial expansion closes on the description of
wars of territorial expansion in different
ant colonies of the same species. Nowadays such a parallel
may seem weird, but at the end of the 19th century it was
fairly natural. In ``The mind and Society'' (sections 155--157)
Pareto
draws several parallels of that kind which are
based on observations by
the entomologist Jean-Henri Fabre. A more comprehensive
study was conducted along this line by the sociologist
Alfred Espinas (Espinas 1876) in his thesis entitled
``On animal societies''.
This was several decades
before Edward Wilson developed sociobiology.


\vskip 3mm

{\bf Acknowledgments} The author would like to express 
his gratitude to Profs. Belal Baaquie,
Pierre Chaunu, Zengru Di, Jack Goldstone, Peter Richmond,
Qinghai Wang,
Jeffrey Williamson and Jinshan Wu who welcomed
the present attempt to give a new life to
analytical history.

\vskip 5mm

{\bf References}

\qparr
Baaquie (B.E.), Roehner (B.M.), Wang (Q.) 2016:
The wage transition in developed countries and its
implications for China. Physica A 470,197-216.

\qparr
Bertillon (L.-A.) 1879: Article ``France'' in the
Dictionnaire Encyclop\'edique des Sciences M\'edicales,
[Encyclopedic Dictionary of the Medical Sciences].
4th series, Vol. 5, p.403-584.\qL
[Available on ``Gallica'', the website of digitized 
publications of the French national library at http://gallica.bnf.fr]

\qparr
Bloch (M.) 1924:
Les rois thaumaturges. 
\'Etude sur le caract\`ere surnaturel attribu\'e \`a la 
puissance royale particuli\`erement en France et en Angleterre.
Strasbourg: Librairie Istra.
Translated into English in 1973 under the title:
The royal touch: sacred monarchy and scrofula in England and France.
London: Routledge and Kegan Paul.

\qparr
Chaunu (P.) 1964: Histoire quantitative ou histoire
s\'erielle.[Quantitative history or serial history].
Cahiers Vilfredo Pareto 2,3,165-176.

\qparr
Chaunu (P.) 1978:
Histoire quantitative, histoire s\'erielle. 
[Quantitative history, serial history,
not yet translated into English].
Paris: Armand Colin.

\qparr
Durkheim (E.) 1897: Le suicide. Paris: Alcan. English
translation: Suicide, a study in sociology. Glencoe, 
Ill.: Free Press.

\qparr
Espinas (A.) 1878, 1935: Des soci\'et\'es animales
[About animal societies].
Thesis, University of Paris. \qL
[In 1879 the thesis was
translated into German and in 1882 into Russian.
It does not seem to have ever been translated into English.
In 1935, it was re-edited by ``Librairie F\'elix Alcan'', Paris.]
\qparr
Goldstone (J.) 1991: Revolution and rebellion in the
early modern world. Berkeley: University of California Press.

\qparr
Lieberson (S.) 1985: Making it count: the improvement
of social research and theory. Berkeley: University of
California Press.

\qparr
Lieberson (S.) 1992: Einstein, Renoir, and Greeley: some
thoughts about evidence in sociology. 1991 Presidential
address. American Sociological Review 57,1-15.

\qparr
Pareto (V.) 1917: Trait\'e de sociologie g\'en\'erale. 
Lausanne: Payot. It was the translation of the
Italian version published in 1916 under
the title: ``Trattato di sociologia generale''.
An English translation appeared in
1935 which was entitled: ``The mind and society''.
New York: Harcourt, Brace and Co.

\qparr
Piketty (T.) 2013: Le capital au XXIe si\`ecle.
Paris: \'Editions du Seuil.\qL
An English translation followed in 2014 with the title:
``Capital in the 21st century''. Harvard University Press.

\qparr
Richmond (P.), Roehner (B.M.) 2013: 
The predictable outcome of house price peaks.
Evolutionary and Institutional Economic Review 9,1,125-139.

\qparr
Richmond (P.), Roehner (B.M.) 2016: 
Effect of marital status on death rates. Part 1: High accuracy
exploration of the Farr–Bertillon effect.
Physica A 450,748-767.

\qparr
Roehner (B.M.) 2001: Hidden collective factors in
speculative trading. A study in analytical economics.
Berlin: Springer-Verlag.

\qparr
Roehner (B.M.) 2002: Pattern and repertoire in history.
Cambridge (Mass.): Harvard University Press.

\qparr
Roehner (B.M.), Rahilly (L.) 2002: Separatism and integration,
A study in analytical history.
Lanham (Maryland): Rowman and Littlefield.

\qparr
Roehner (B.M.), Rahilly (L.) 2016: Separatism and disintegration.
A comparative investigation. \qL
[Available at: http://www.lpthe.jussieu.fr/~roehner/separatism.pdf]

\qparr
Roehner (B.M.) 2003a: Predictions in analytical history.
Historically speaking 5,1,37-39.

\qparr
Roehner (B.M.) 2003b: Patterns of speculation in real estate
and stocks. in Nikkei Workshop on Econophysics (2003). 
Berlin: Springer Verlag. 

\qparr
Roehner (B.M.) 2006:
Real estate price peaks: a comparative perspective.
Evolutionary and Institutional Economics Review 2,2,167-182.

\qparr
Tilly (C.) 1981: As sociology meets history. New York: Russel
Sage.

\qparr
Tilly (C.) 1995, 1998: Popular Contention in Great Britain, 1758–1834.
Cambridge (Ma): Harvard University Press

\qparr
Tracy (M.)  2014: Capital: A hit that was, wasn't, then was again.
How the French tome
has rocked the tiny Harvard University Press. 
New Republic, 24 April.

\end{document}